# Applying News and Media Sentiment Analysis for Generating Forex Trading Signals

**O.F. Olaiyapo**
Emory University, Atlanta, USA

**ABSTRACT**

The **objective** of this research is to examine how sentiment analysis can be employed to generate trading signals for the Foreign Exchange (Forex) market. The author assessed sentiment in social media posts and news articles pertaining to the United States Dollar (USD) using a combination of **methods**: lexicon-based analysis and the Naive Bayes machine learning algorithm. The **findings** indicate that sentiment analysis proves valuable in forecasting market movements and devising trading signals. Notably, its effectiveness is consistent across different market conditions. The author **concludes** that by analyzing sentiment expressed in news and social media, traders can glean insights into prevailing market sentiments towards the USD and other pertinent countries, thereby aiding trading decision-making. This study underscores the importance of weaving sentiment analysis into trading strategies as a pivotal tool for predicting market dynamics.

# Применение анализа тональности настроений в средствах массовой информации и социальных сетях для генерации торговых сигналов на рынке Форекс

**О.Ф. Олайапо**
Университет Эмори, Атланта, США

**АННОТАЦИЯ**

**Цель** данного исследования — изучить возможность анализа настроений для генерации торговых сигналов на валютном рынке (Forex). Автор оценивал тональность настроений в сообщениях социальных сетей и статьях средств массовой информации, касающихся доллара США, используя комбинацию **методов**: анализа на основе лексики и алгоритма наивного байесовского классификатора. Полученные **результаты** свидетельствуют о том, что анализ настроений является ценным инструментом для прогнозирования движения рынка и разработки торговых сигналов. Отмечается, что его эффективность неизменна в различных рыночных условиях. Автор делает **вывод**, что, анализируя настроения, выраженные в новостях и социальных сетях, трейдеры могут получить представление о преобладающих на рынке отношениях к доллару США и валютам других стран, тем самым способствуя принятию торговых решений. Данное исследование подчеркивает важность включения анализа настроений в торговые стратегии в качестве ключевого инструмента для прогнозирования динамики рынка.
*Ключевые слова:* рынок Форекс; настроения; торговые сигналы; рынок иностранных валют; валюта; социальные сети; наивный байесовский классификатор; машинное обучение

## 1. Introduction

The Forex market, also known as the Foreign Exchange market, is a marketplace where currencies are traded in pairs [1]. Traders engage in buying and selling currencies, aiming to profit from changes in their exchange rates. This market is highly dynamic and volatile, influenced by factors such as news, social media posts and political events. To stay ahead of the game, traders need to stay updated on the news and events that can impact currency exchange rates [2]. Having the ability to accurately predict when to buy or sell prices is a tool for traders and investors.

In this paper, we aim to explore the potential of sentiment analysis for generating trading signals. While studies have been conducted on using news data to predict stock prices, there has been limited research on applying sentiment analysis to trading.

Sentiment analysis has emerged as a tool for understanding market sentiments and forecasting currency pair movements within the market. Our main objective is to utilize sentiment analysis techniques on both news data and social media information to generate trading signals.

## 2. Literature review

The prediction of currency exchange rates has been a puzzle and an area of interest for researchers, economists and financial policymakers. Sentiment analysis, which has historically played a role in fields like media analysis, product reviews and financial markets, has now gained prominence as a tool. In the realm of trading, sentiment analysis has become an element for predicting market movements. This literature review thoroughly explores research that utilizes sentiment analysis in trading.

One pioneering study conducted in 2011 took an approach by combining Google Trends data with sentiment analysis of news articles to forecast stock market movements. Their methodology showed an impact on predicting stock market trends, laying the groundwork for researchers [3].

Digging deeper into the practicality of sentiment analysis, the research conducted by Chen et al. (2014) deserves attention [4]. Focusing on analyzing sentiments in news articles, they employed a support vector machine (SVM) classifier to identify whether sentiments were neutral, positive or negative. Notably, their findings emphasized the superiority of their approach compared to baseline methods in predicting stock returns.

Combining long short-term memory (LSTM) and convolutional neural network (CNN), for Social Media Sentiment. Shifting towards methods and datasets, the work carried out by Zhang and colleagues in 2019 [5] stands out. They decided to explore the sea of Twitter[1, 2] data using a combination of advanced LSTM and CNN architectures. Their main goal was to classify the sentiments expressed in Twitter posts into negative categories. The results they obtained were quite compelling, indicating that sentiment analysis on real time platforms, such as Twitter, has an impact when it comes to predicting forex exchange rates.

Most studies adopt a perspective by focusing on well-known currency pairs without exploring the intricacies and potential trading opportunities offered by lesser-known pairs. There is also a lack of research on the impact of sentiment analysis on different currency pairs. There is a need for more research on the integration of sentiment analysis with other technical indicators to improve Forex trading strategies.

## 3. Problem formulation and research questions

This paper aims to utilize sentiment analysis to generate Forex trading signals by using both news articles and social media data. Trading forex involves buying and selling currencies with the aim of making profit based on exchange rate fluctuations; it also involves buying and selling currency pairs based on their relative value. Traders employ tools and strategies to analyze market trends and make informed trading decisions. A crucial factor influencing market trends is the impact of news and events on the economies of countries whose currencies are being traded.

Sentiment analysis plays a role in examining text data from news articles and social media posts to uncover sentiments [6–8]. It entails identifying and categorizing the polarity of expressed sentiment in text as positive, negative or neutral.

---

[1] Twitter — social network. Current name X.

[2] Twitter is blocked in Russia.

Table 1
*Research Questions and Objectives*

| Research Question | Objective |
|---|---|
| Can sentiment analysis of news data be used to generate Forex trading signals? | To collate Forex-related news data |
| What are the most effective sentiment analysis techniques for generating Forex trading signals? | To perform sentiment analysis on news data using two techniques which are lexicon-based analysis and Naive Bayes algorithm |
| How accurate are the Forex trading signals generated through sentiment analysis? | To evaluate the accuracy of the signals generated using sentiment analysis |

*Source:* Developed by the author.

To develop a system that generates trading signals based on sentiment analysis results, we will utilize both lexicon based and machine learning based approaches. The lexicon-based approach involves employing the VADER (Valence Aware Dictionary and Sentiment Reasoner) tool to determine the polarity of news articles. On the other hand, the machine learning approach utilizes the Naive Bayes algorithm for sentiment classification of these news articles [9]. Additionally, incorporating indicators as confirmation signals to enhance the accuracy of trading signals will be explored.

To accomplish this objective, a dataset comprising news articles pertaining to USD and other major currencies, in the market will be utilized. The dataset contains text data, as well as metadata such as the publication date and the source of the news article. The dataset was collected from reputable news sources such as Reuters, Bloomberg, and Twitter posts.

The main research questions to address in this paper are presented in *Table 1.*

By answering these research questions, this paper aims to provide insights into the potential of sentiment analysis in Forex trading and its impact on market trends.

## 4. Technical design

The technical design of this sentiment analysis based on a forex trading system involves different components, which include data pre-processing, sentiment analysis, and trading signal generation. In this section, a detailed description of the methodology and the techniques used for the components is provided.

### 4.1. Data pre-processing

News articles were collected from financial news websites and social media platforms that are related to the United States Dollar (USD). The USD is one of the major currencies used in the Forex market, and it is considered the world's most dominant reserve currency. It is widely accepted in international transactions. The data were preprocessed to remove noise and irrelevant information, such as stop words and punctuation. Text normalization was also performed; this involves converting all text to lowercase and removing any special characters and symbols [10, 11].

### 4.2. Sentiment analysis techniques

Two main techniques used include Lexicon-based analysis and Naïve Bayes.

### 4.2.1. Lexicon-based analysis

After the words were pre-processed, the VADER (Valence Aware Dictionary and Sentiment Reasoner) tool was utilized. VADER uses a predefined sentiment lexicon to assign a polarity score to each text document in the dataset. This contains a list of words and their associated neutral, positive, and negative scores. VADER assigns a score between –1 and +1 to each text document, where –1 indicates highly negative sentiment, and +1 indicates highly positive sentiment, and 0 indicates neutral sentiment.

To obtain an overall sentiment score for each currency pair, the polarity scores of all the news articles related to that currency pair were aggregated. The aggregation was done using a weighted average, where the weight of each polarity score was determined based on the relevance of the news article to the currency pair.

$$Sentiment\ Score = \frac{\sum(Polarity\ Score \times Weight)}{\sum(Weight)}, \quad (1)$$

where Polarity score is the sentiment polarity score assigned by VADER to the news article or social media post, and Weight is the relevance weight assigned to the news article.

$$Polarity\ Score = \frac{\sum_{i=1}^{n} Valence_i}{\sqrt{\sum_{i=1}^{n} Valence_i^2 + \alpha}}. \quad (2)$$

The above is used by VADER, where valence scores are the set of values assigned to each word in the sentiment lexicon, indicating its positivity or negativity. Alpha, in this case, is a smoothing factor to normalize the score. This algorithm was selected because, unlike other algorithms, it doesn't solely focus on words; it also considers the context in which they are used. This distinction enables it to effectively discern the sentiment behind nuanced statements commonly found on social media platforms. Furthermore, it is adeptly tuned to grasp the sentiment conveyed by slang, acronyms, and prevalent internet terms. When juxtaposed with models like deep learning approaches, VADER achieves a harmonious balance between computational efficiency and accuracy.

### 4.2.2. Naïve Bayes

Naïve Bayes is a machine learning algorithm that can be used for classification. It is also a probabilistic algorithm that works on the assumption that each feature is independent of all other features. To use this for sentiment analysis, the model needs to be trained using a dataset of news articles and social media posts. In this case, they were labeled as either neutral, positive, or negative.

After training the model, it was used to classify the sentiment of the news articles and social media posts that are related to USD. The model then computed the probability that a given text in the news article or social media post belonged to each sentiment class (positive, negative, or neutral) based on the frequency of words in the text that appeared in each class of the training data.

The application of Naïve Bayes theorem to a class $y$ with three possible outcome (neutral, positive, or negative) and a dependent vector $x = (x_1, \ldots, x_n)$ is expressed as:

$$P(y|x) = \frac{P(x|y)P(y)}{P(x)}, \quad (3)$$

where:
- $P(y|x)$: The probability that a given piece of text falls under sentiment class $y$ based on the word frequencies represented by vector $x$.
- $P(x|y)$: Likelihood of encountering the word frequencies in vector $x$ in texts classified under sentiment class $y$.
- $P(y)$: Base probability of a text being categorized under sentiment class $y$, without considering its word frequencies.
- $P(x)$: Likelihood of the word frequencies denoted by a vector appearing across a sentiment classification.

If one assumes that all features are independent, this relationship can be further simplified to:

$$P(x_i \mid y, x_i, \ldots, x_{i-1}, x_{i+1}, \ldots, x_{ni}) = P(x_i|y). \quad (4)$$

- $P(x_i \mid y)$: Likelihood of a specific word $x_i$ appearing in texts of sentiment class $y$, given the assumption that words appear independently.

Using the above equation, we deduce:

$$P(y \mid x) = P(y) \prod_{i=1}^{n} P(x_i|y). \quad (5)$$

- $\prod_{i=1}^{n} P(x_i \mid y)$: This term sums up the likelihood of all words in vector $x$ showing up in texts under sentiment class $y$.

Given a constant $P(x)$ for the input, classification is determined by:

$$y = \text{argmax}_y P(y) \prod_{i=1}^{n} P(x_i | y). \quad (6)$$

- $\text{argmax}_y$: This function pinpoints the sentiment class $y$ which has the highest overall probability when considering the words in vector $x$.

For estimating $P(y)$ and $P(x_i | y)$, a Gaussian assumption is leverage on the dependent variable through the Maximum A Posteriori (MAP):

$$P(x_i | y) = \frac{1}{\sqrt{2\pi\sigma_y^2}} \exp^{-\frac{(x_i - \mu_y)^2}{2\sigma_y^2}}. \quad (7)$$

A rejection model variant for classification is used here; for a classification to be accepted, the probability needs to be greater than a specific threshold, or projection. If this threshold is not met, the mode will reject classifying a given sample. The Naïve Bayes classifier is useful for sentiment analysis because its computational complexity is light weight, making it resilient to over-fitting and suitable for use with large datasets from news articles and social media posts [5].

To further illustrate the Naïve Bayes procedure. Here is an example of how this model works. Suppose a news article is titled "Federal Reserve announces interest rate hike, boosting dollar". After the noise is removed from this news article. The probability is calculated that the words in the news data appear in each sentiment class, which is neutral, positive, or negative based on the training data. The word "boosting", for example, appears to be in positive texts rather than in negative or neutral texts.

The probability that the text belongs to one of the sentiment classes based on the probabilities of the individual words is then calculated [8]. By using this procedure on a large dataset of news articles and social media posts, the trading signals were generated based on the prevailing sentiment towards USD.

This approach was chosen because of its simplicity and rapid performance, making it ideally suited for generating real-time Forex trading signals. While other machine learning methods, such as support vector machines (SVM) or decision trees, offer nuanced decision boundaries, the probabilistic nature of Naive Bayes facilitates sentiment estimation based on the provided features. Its inherent capability to manage feature spaces, which are common in text data, grants it a distinct advantage.

**4.3. Trading signal generation process**

After the sentiment analysis scores for the news articles and social media data are obtained, the next step is to use the scores to generate Forex trading signals based on the sentiment analysis results. Forex trading signals are indications that tell Forex traders when to buy, sell or hold a particular currency pair based on market analysis.

To generate this Forex trading signal, the sentiment analysis results were combined with technical analysis indicators. Technical analysis is a forex market evaluation process that assists traders in predicting the next direction of currency pair values based on past price movements and chart patterns. According to technical traders, markets usually move in predictable patterns that may be observed and identified on trading charts.

There are numerous technical analysis indicators that are used in Forex trading that can be analyzed using data mining techniques to generate trading signals. These indicators are used to confirm the signals generated by the sentiment analysis. Some of the technical indicators include moving average convergence divergence (MACD), which helps Forex traders identify emerging price trends, upward or downward. Bollinger Bands, which is used to lay out trend lines two standard deviations away from the simple moving average price of a financial instrument. Stochastic Oscillator helps to compare a currency pair's closing price to its price range over a specific period.

The moving average (MA), which is the average price of a currency pair over a set period, was used. The relative strength index (RSI), which measures the strength and weakness of a currency pair's price action, was also used. It ranges from zero to 100, and it identifies when a currency is overbought or oversold in market conditions. The selection of MSI and MA indicators stems from their established efficacy in pinpointing price momentum and trends. These indicators enhance clarity and furnish decisive signals when evaluating sentiment within Forex trading.

When a sentiment analysis score and these two technical analysis indicators align, then a trading signal is generated. For example, if the sentiment analysis indicates a positive score and the technical analysis indicator also confirms and indicates a bullish trend, then a "buy" trade signal is generated.

The sentiment analysis result was positive, and the currency price was above its 50-day moving average (MA) and the relative strength index (RSI) was above 50, we generated a stronger buy signal than if only the sentiment analysis result was positive. On the other hand, if the sentiment analysis indicates a negative score and the technical analysis indicators also indicate a bearish trend, then a "sell" trade signal is generated.

If the sentiment analysis and the technical indicators do not align, then no signal will be generated, and traders would have to wait and exercise caution. Forex trading involves risk, so traders should always exercise caution when the sentiment analysis and the technical indicators do not align.

## 5. Experimental analysis

This section shows the experimental setup and methodology used in this research, followed by a presentation of the results and an analysis of the findings.

### 5.1. Experimental setup and methodology

The models used for the sentiment analysis were implemented in the Python programming language using the scikit-learn library. The dataset consists of news articles from financial news websites and social media posts that are related to the USD. After the dataset was pre-processed, i.e., removing noise such as irrelevant information as mentions, links, and hashtags. The training set was used to train the sentiment analysis models and the testing set was used to evaluate the performance of the models.

Two models of sentiment analysis were used for this research, which include lexicon-based analysis and Naïve Bayes. The lexicon-based models classify news articles and social media posts using pre-built sentiment. While the Naïve Bayes model was trained using Gaussian assumption on the dependent variable to estimate $P(y)$ and $P(x_i|y)$.

The scores generated by the sentiment analysis models were used to generate Forex trading signals. To confirm the accuracy of the generated signals, we use two technical indicators: the moving average (MA) and the relative strength index (RSI). The RSI indicator measures the convergence and divergence of two moving averages and identifies potential trend changes. The MA identifies trends and support, and resistance levels.

### 5.2. Detailed experimental specifics
#### 5.2.1. Duration and long-term trends

The experiment was based on historical data from April 1st, 2023, to July 26th, 2023. During this timeframe, there was notable market volatility. The US dollar witnessed a strengthening trend against various currencies due to several factors. These included the Federal Reserve's decision to aggressively raise interest rates to tackle inflation, the ongoing conflict in Ukraine and its impact on the global economy, as well as China's economic slowdown. Among the currencies that were hit the hardest were the euro (EUR) and pound sterling (GBP), experiencing declines of 10% and 15% against the US dollar during this timeframe, respectively. Additionally, the Japanese yen (JPY) also experienced a depreciation of around 17% compared to the USD.

#### 5.2.2. Timeframe

For the purpose of this research, the H4 timeframe was chosen for analysis. This timeframe strikes a balance between the short-term fluctuations observed in M15 (15-minute) or H1 (1-hour) timeframes and the longer-term trends seen in weekly timeframes. Since news articles and social media posts used for sentiment analysis may not capture immediate market reactions, analyzing the H4 timeframe provides a broader window to observe how sentiment translates into price action.

#### 5.2.3. Volatility and slippage

Between April 1st and July 26th, 2023, the Forex market experienced significant volatility. Due to this heightened volatility, a discrepancy was observed between the price at which orders were placed and the price at which they were ex-

Table 2
*Precision, recall and F1 score*

| Model | Accuracy | Precision | Recall | F1 Score |
|---|---|---|---|---|
| Naïve Bayes | 85% | 0.87 | 0.85 | 0.86 |
| Lexicon-based | 70% | 0.72 | 0.70 | 0.71 |

*Source:* Developed by the author.

Table 3
*EUR/USD historical data*

| Date | Price | Open | High | Low | Vol. | Change, % |
|---|---|---|---|---|---|---|
| 04/01/2023 | 1.0987 | 1.0847 | 1.1076 | 1.0789 | | +1.37% |
| 03/01/2023 | 1.0839 | 1.0576 | 1.0931 | 1.0516 | | +2.49% |
| 02/01/2023 | 1.0576 | 1.0864 | 1.1034 | 1.0532 | | −2.63% |
| 01/01/2023 | 1.0862 | 1.0702 | 1.0930 | 1.0483 | | +1.50% |
| 12/01/2022 | 1.0702 | 1.0407 | 1.0737 | 1.0393 | | +2.85% |
| 11/01/2022 | 1.0405 | 0.9882 | 1.0497 | 0.9729 | | +5.28% |
| 10/01/2022 | 0.9883 | 0.9800 | 1.0094 | 0.9632 | | +0.86% |
| 09/01/2022 | 0.9799 | 1.0055 | 1.0199 | 0.9535 | | −2.57% |
| 08/01/2022 | 1.0057 | 1.0218 | 1.0369 | 0.9900 | | −1.58% |
| 07/01/2022 | 1.0218 | 1.0481 | 1.0487 | 0.9952 | | −2.52% |
| 06/01/2022 | 1.0482 | 1.0734 | 1.0775 | 1.0359 | | −2.34% |
| 05/01/2022 | 1.0733 | 1.0550 | 1.0788 | 1.0350 | | +1.82% |
| 04/01/2022 | 1.0541 | 1.1065 | 1.1077 | 1.0471 | | −4.74% |
| 03/01/2022 | 1.1065 | 1.1218 | 1.1234 | 1.0805 | | −1.37% |
| 02/01/2022 | 1.1219 | 1.1234 | 1.1496 | 1.1106 | | −0.12% |
| 01/01/2022 | 1.1233 | 1.1365 | 1.1484 | 1.1121 | | −1.19% |
| 12/01/2021 | 1.1368 | 1.1339 | 1.1387 | 1.1221 | | +0.28% |
| 11/01/2021 | 1.1336 | 1.1559 | 1.1617 | 1.1185 | | −1.95% |
| 10/01/2021 | 1.1561 | 1.1582 | 1.1693 | 1.1524 | | −0.17% |
| 09/01/2021 | 1.1581 | 1.1810 | 1.1910 | 1.1562 | | −1.91% |
| 08/01/2021 | 1.1807 | 1.1864 | 1.1901 | 1.1664 | | −0.53% |
| 07/01/2021 | 1.1870 | 1.1859 | 1.1910 | 1.1752 | | +0.13% |
| 06/01/2021 | 1.1855 | 1.2227 | 1.2255 | 1.1845 | | −3.03% |
| 05/01/2021 | 1.225 | 1.2031 | 1.2267 | 1.1985 | | +1.72% |

*Note:* Highest: 1.2267; Lowest 0.9535; Difference: 0.2731; Average: 1.0967; Change, %: −8.5788.

*Source:* Adopted by the author from Investing.com

Table 4
*USD/CAD historical data*

| Date | Price | Open | High | Low | Vol. | Change, % |
| --- | --- | --- | --- | --- | --- | --- |
| 04/23/2023 | 1.3541 | 1.3542 | 1.3546 | 1.3534 | 3.20K | +0.01% |
| 04/21/2023 | 1.3539 | 1.3475, | 1.3564 | 1.3471 | 55.83K | +0.48% |
| 04/20/2023 | 1.3474 | 1.3460 | 1.3491 | 1.3448 | 57.85K | +0.10% |
| 04/19/2023 | 1.3460 | 1.3389 | 1.3469 | 1.3384 | 57.74K | 0.55% |
| 04/18/2023 | 1.3387 | 1.3395, | 1.3400 | 1.3359 | 57.14K | −0.04% |
| 04/17/2023 | 1.3392 | 1.3362 | 1.3421 | 1.3341 | 55.30K | +0.23% |
| 04/14/2023 | 1.3361 | 1.3338 | 1.3397 | 1.3302 | 63.92K | +0.20% |
| 04/13/2023 | 1.3334 | 1.3441 | 1.3449 | 1.3333 | 58.92K | −0.77% |
| 04/12/2023 | 1.3438 | 1.3468 | 1.3491 | 1.3427 | 57.14K | −0.21% |
| 04/11/2023 | 1.3466 | 1.3510 | 1.3518 | 1.3462 | 54.91K | −0.30% |
| 04/10/2023 | 1.3506 | 1.3523 | 1.3555 | 1.3484 | 53.44K | −0.01% |
| 04/07/2023 | 1.3507 | 1.3492 | 1.3532 | 1.3483 | 42.11K | +0.11% |
| 04/06/2023 | 1.3492 | 1.3457 | 1.3507 | 1.3447 | 55.31K | +0.25% |
| 04/05/2023 | 1.3458, | 1.3442 | 1.3484 | 1.3426 | 55.04K | +0.14% |
| 04/04/2023 | 1.3439 | 1.3438 | 1.3469 | 1.3406 | 57.97K | +0.03% |
| 04/03/2023 | 1.3435, | 1.3518 | 1.3537, | 1.3411 | 60.62K | −0.59% |
| 03/31/2023 | 1.3515 | 1.3523 | 1.3566 | 1.3507 | 59.61K | −0.04% |
| 03/30/2023 | 1.3520 | 1.3558 | 1.3583 | 1.3515 | 57.38K | −0.26% |
| 03/29/2023 | 1.3555 | 1.3602 | 1.3618 | 1.3556 | 56.09K | −0.32% |
| 03/28/2023 | 1.3598 | 1.3663 | 1.3697 | 1.3591 | 59.52K | −0.45% |
| 03/27/2023 | 1.3660 | 1.3741 | 1.3747 | 1.3649 | 63.27K | −0.61% |
| 03/24/2023 | 1.3744 | 1.3717 | 1.3806 | 1.3707 | 70.30K | +0.23% |
| 03/23/2023 | 1.3713 | 1.3728 | 1.3726 | 1.3630 | 68.81K | −0.10% |

*Note:* Highest: 1.3806; Lowest: 1.3302; Difference: 0.0504; Average: 1.3501; Change %: −1.3550.

*Source:* Adopted by the author from Investing.com

ecuted — a phenomenon commonly referred to as slippage.

This slippage was especially pronounced in timeframes like M1 and M5, where algorithmic trading systems respond instantly to information. However, in the H4 timeframe these algorithms react with a cushion against news events and sentiment scores, helping to mitigate the impact of slippage.

### 6. Results and analysis

The generated forex trading signals were evaluated using historical forex data. With confirmation from the MA and RSI, the signals generated by the Naïve Bayes model showed that a profit of over 12% was gained over the testing period. While the lexicon-based models using confirmation from the RSI and MA indicators generated a profit of 5% over the testing period.

Naïve Bayes is the most effective in analyzing sentiment in this research project based on the evaluation metrics. The accuracy of the Naïve Bayes model indicates that it correctly classified 85% of the news articles and social media posts as either neutral, negative, or positive sentiments. The lexicon-based model

Table 5
*GBP/USD historical data*

| Date | Price | Open | High | Low | Vol. | Change, % |
| --- | --- | --- | --- | --- | --- | --- |
| 04/01/2023 | 1.2445 | 1.2332 | 1.2547 | 1.2275 | | +0.91% |
| 03/01/2023 | 1.2333 | 1.2021 | 1.2424 | 1.1803 | | +2.60% |
| 02/01/2023 | 1.2020 | 1.2319 | 1.2401 | 1.1914 | | −2.44% |
| 01/01/2023 | 1.2320 | 1.2098 | 1.2450 | 1.1840 | | +1.84% |
| 12/01/2022 | 1.2097 | 1.2056 | 1.2448 | 1.1992 | | +0.34% |
| 11/01/2022 | 1.2056 | 1.1471 | 1.2155 | 1.1148 | | +5.12% |
| 10/01/2022 | 1.1469 | 1.1161 | 1.1647 | 1.0924 | | +2.77% |
| 09/01/2022 | 1.1160 | 1.1623 | 1.1740 | 1.0384 | | −3.98% |
| 08/01/2022 | 1.1622 | 1.2165 | 1.2294 | 1.1598 | | −4.47% |
| 07/01/2022 | 1.2166 | 1.2177 | 1.2247 | 1.1759 | | −0.07% |
| 06/01/2022 | 1.2175 | 1.2604 | 1.2618 | 1.1933 | | −3.37% |
| 05/01/2022 | 1.2600 | 1.2587 | 1.2667 | 1.2155 | | +0.23% |
| 04/01/2022 | 1.2571 | 1.3136 | 1.3168 | 1.2410 | | −4.28% |
| 03/01/2022 | 1.3133 | 1.3421 | 1.3439 | 1.2999 | | −2.13% |
| 02/01/2022 | 1.3419 | 1.3447 | 1.3645 | 1.3271 | | −0.19% |
| 01/01/2022 | 1.3445 | 1.3531 | 1.3750 | 1.3357 | | −0.62% |
| 12/01/2021 | 1.3529 | 1.3299 | 1.3551 | 1.3367 | | +1.76% |
| 11/01/2021 | 1.3295 | 1.3683 | 1.3699 | 1.3195 | | −2.89% |
| 10/01/2021 | 1.3691 | 1.3472 | 1.3836 | 1.3434 | | +1.63% |
| 09/01/2021 | 1.3472 | 1.3753 | 1.3914 | 1.3412 | | −2.05% |
| 08/01/2021 | 1.3754 | 1.3901 | 1.3958 | 1.3601 | | −1.06% |
| 07/01/2021 | 1.3901 | 1.3829 | 1.3985 | 1.3571 | | +0.54% |
| 06/01/2021 | 1.3827 | 1.4213 | 1.4250 | 1.3786 | | −2.69% |
| 05/01/2021 | 1.4209 | 1.3827 | 1.4234 | 1.3800 | | +2.86% |

*Note:* Highest: 1.4250; Lowest: 1.0384; Difference: 0.3865; Average: 1.2780; Change %: −9.9102.

*Source:* Adopted by the author from Investing.com

has lower accuracy, which suggests that it may not be as effective as Naïve Bayes in identifying the correct sentiment in Forex trading (*Table 2*).

The lexicon-based model has a precision score of 0.72, a recall of 0.70 and an F1 score of 0.71. The Naïve Bayes model has a precision score of 0.87, recall of 0.85 and an F1 score of 0.86. These results suggest that 87% of the positive classifications made by the Naïve Bayes model were indeed correct when it classified news articles or social media posts as either neutral, positive, or negative. When the model encounters a neutral, positive, or negative sentiment, 85% of the time the recall score correctly identified it. The F1 score shows that the Naïve Bayes model has a good balance between precision and recall.

In terms of using technical analysis indicators for confirmation, the analysis showed that the moving average (MA) and relative strength index (RSI) are effective in confirming the trading signal models generated by the sentiment analysis model. The sentiment analysis can generate a positive score, and then the MA and RSI confirm the buy signal, resulting in a profitable trade. Similarly, the models can also generate a negative score, and the MA and RSI confirm the sell signal, which results in a profitable trade (*Tables 3–5*).

**Limitations and potential improvements**

Sentiment analysis can sometimes be biased, and this system relies on it. One potential improvement would be to use more advanced sentiment analysis, such as deep learning approaches, which can be more effective in capturing nuances in language.

This paper also shows that sentiment analysis is good for signal generation in the Forex market and can improve traders' performance.

### 7. Conclusion and future work

The significance of this research lies in its potential to enhance the quality of generating profitable Forex trading signals and helping Forex traders in making decisions. This research project has investigated the potential of sentiment analysis to generate Forex trading signals. This project proposed and implemented a methodology based on lexicon-based analysis and Naïve Bayes to assign a neutral, positive, or negative score to the sentiment of news articles and social media posts related to United States Dollars (USD), and then combine technical analysis indicators to generate a signal.

As for future work, there are several areas where further research and investigation can be conducted. Since there are different factors that affect the Forex market other than financial news articles or social media posts, advanced sentiment analysis techniques such as deep learning and neural networks can be used on various factors, which include political events, economic data releases, natural disaster news, among others, to further provide traders with useful insights to generate profitable Forex trading signals.

Also, exploring the application of sentiment analysis in other financial markets beyond Forex, such as stocks or commodities and crypto trading, could also be an interesting avenue for future research.

## ABOUT THE AUTHOR / ИНФОРМАЦИЯ ОБ АВТОРЕ

***Oluwafemi F. Olaiyapo*** — Graduate Student, Department of Mathematics, Emory University, Atlanta, USA
***Олувафеми Ф. Олайапо*** — аспирант кафедры математики, Университет Эмори, Атланта, США
https://orcid.org/0009-0000-8147-0204
oluwafemiolaiyapo@gmail.com